# Using Artificial Intelligence to Accelerate Collective Intelligence

Policy Synth and Smarter Crowdsourcing

Róbert Bjarnason

President and Chief Technology Officer, Citizens Foundation, robert@citizens.is

Dane Gambrell

Research Fellow, The Governance Lab, dane@thegovlab.org

Joshua Lanthier-Welch

Executive Director, Citizens Foundation America, josh@citizens.is

In an era characterized by rapid societal changes and complex challenges, institutions' traditional methods of problem-solving in the public sector are increasingly proving inadequate. In this study, we present an innovative and effective model for how institutions can use artificial intelligence to enable groups of people to generate effective solutions to urgent problems more efficiently. We describe a proven collective intelligence method, called Smarter Crowdsourcing, which is designed to channel the collective intelligence of those with expertise about a problem into actionable solutions through crowdsourcing. Then we introduce Policy Synth, an innovative toolkit which leverages AI to make the Smarter Crowdsourcing problem-solving approach both more scalable, more effective and more efficient. Policy Synth is crafted using a human-centric approach, recognizing that AI is a tool to enhance human intelligence and creativity, not replace it. Based on a real-world case study comparing the results of expert crowdsourcing alone with expert sourcing supported by Policy Synth AI agents, we conclude that Smarter Crowdsourcing with Policy Synth presents an effective model for integrating the collective wisdom of human experts and the computational power of AI to enhance and scale up public problem-solving processes.

The potential for artificial intelligence to enhance the performance of groups of people has been a topic of great interest among scholars of collective intelligence. Though many AI toolkits exist, they too often are not fitted to the needs of institutions and policymakers. While many existing approaches view AI as a tool to make crowdsourcing and deliberative processes better and more efficient, Policy Synth goes a step further, recognizing that AI can also be used to synthesize the findings from engagements together with research to develop evidence-based solutions and policies. This study contributes significantly to the fields of collective intelligence, public problem-solving, and AI. The study offers practical tools and insights for institutions looking to engage communities effectively in addressing urgent societal challenges.

## 1 INTRODUCTION

In this study, we introduce a set of practical tools and methods that institutions can use to engage groups in solving public problems using a combination of artificial intelligence (AI) and collective intelligence (CI).

First, we describe Smarter Crowdsourcing – a proven approach that institutions are already using to engage the collective intelligence of human experts in efforts to solve public problems. Second, we introduce a new toolkit, called Policy Synth, which leverages artificial intelligence techniques to automate and scale up the Smarter Crowdsourcing method. By integrating AI capabilities into this process, Policy Synth aims to make the Smarter Crowdsourcing method more efficient, effective, and scalable, enabling institutions to develop solutions more rapidly than would be possible with human intelligence alone.

Finally, we describe a case study comparing solutions to a complex policy problem generated using artificial intelligence to those developed through expert crowdsourcing. The case study demonstrates how generative AI can make participatory problem-solving with human participants more scalable and more effective because the tools make it possible to translate ideas into implementable proposals more quickly than human participation alone.

This study contributes to the body of research around collective intelligence for public problem-solving [Gambrell and Noveck 2020] as well as the use of AI for public good.

## 2 THE NEED FOR RAPID AND EFFICIENT PUBLIC PROBLEM-SOLVING

From climate change to economic instability and social inequality, to public health crises, policymakers and institutions face a range of urgent challenges. The ability to quickly identify, understand, and address these problems is crucial for maintaining societal stability and fostering sustainable progress. Yet, the mechanisms that institutions use to understand and address these challenges too often fail to develop and implement effective solutions at scale [Noveck 2021] [Mulgan 2017].

The rapid pace of societal advancement and technological change is deepening this challenge. Public institutions now work in increasingly complex and uncertain environments, with the world's growing interconnectivity making our challenges larger and more complex. The failure to develop effective solutions and create meaningful opportunities for communities to participate in solving problems may be contributing to a lack of trust in government [Kumagai and Iorio 2020]. In countries around the world, there is a broadly-held belief among members of the public that participating in elections – the most direct opportunity many residents have to influence public policy – simply does not matter [Van Reybrouck 2018]. There is a risk that the democratic backsliding that has been observed in the United States and other democracies will accelerate as people continue to see their governments as ineffective at solving problems that matter to them [Light 2014].

Fortunately, institutions around the world are experimenting with more participatory methods to solve problems, and in particular, to meaningfully incorporate the collective intelligence of communities and subject matter experts into their problem-solving processes. These include experiments that engage groups in lawmaking and policymaking at all levels of government [Noveck, et al. 2020] as well as efforts to institutionalize large-scale collaboration, crowdsourcing, and co-creation projects as part of institutions' formal problem-solving and decisionmaking processes [Ryan, et al. 2020]. As we will show, the rapid advance in technology – most notably, the proliferation of generative AI tools – presents great opportunities to further scale these efforts.

In the following section, we elaborate on one collective intelligence method – Smarter Crowdsourcing – that we have used to aid institutions in addressing problems and discuss the challenges to efficiently using collective intelligence for problem solving and then address how generative AI is helping to make this process more efficient and effective.

# 3 SMARTER CROWDSOURCING: RAPIDLY CROWDSOURCING EXPERTISE

Smarter Crowdsourcing is a problem-solving method that leverages the collective intelligence of diverse experts to identify solutions to urgent public problems in support of governments and philanthropic organizations [Dinesh 2023]. The process marries the diversity and agility of crowdsourcing with curation, to find and bring together those with relevant know-how and practical experience, in a format designed to produce the most innovative solutions to challenging public problems.

Smarter Crowdsourcing was developed and tested for the first time in 2015 as part of an effort to advise the government of Quito, Ecuador about how to prepare for the eruption of the Cotopaxi Volcano. After this pilot, the methodology was further refined through a series of projects with other institutions. Following the Cotopaxi use case, the Smarter Crowdsourcing has been used seven additional times to support governments and philanthropic organizations:

- Zika (2016): Explored ways to address the causes of Zika virus and other mosquito-borne diseases in support of the Inter-American Development Bank and partner governments in four countries in Latin America.
- Anti-Corruption (2017): Worked with the Mexican government and civil society institutions to explore solutions to the problem of corruption in Mexico.
- Coronavirus (2020): Supported the Inter-American Development Bank and its partner governments in Latin America and the Caribbean to source solutions to problems arising from the COVID-19 pandemic.
- Education (2020): Developed solutions to address the challenge of equitably identifying and measuring the non-academic skills young people need to succeed to inform the Walton Family Foundation's funding strategy.
- Experiential Learning (2021): Explored innovations in equity-centered hands-on learning initiatives in support of Northeastern University.
- Modernization of Congress (2021): Developed recommendations on evidence-based lawmaking in support of the US House Select Committee on the Modernization of Congress.
- Countering Election Subversion (2023): Developed recommendations for solutions philanthropy could support in order to counter efforts to undermine or cast doubt upon the integrity of U.S. elections in support of Democracy Fund Voice.

In each case, the process resulted in a set of practical and innovative solutions tailored to the partner institution's needs and priorities, in terms of format as well as content. For example, the Smarter Crowdsourcing: Education initiative described below generated 22 specific funding opportunities designed to have real-world positive impact, each one with a concrete action plan. These recommendations guided the strategic direction of the Walton Family Foundation's grantmaking portfolio in the area of non-academic skills assessment.

Smarter Crowdsourcing was developed well before the use of generative AI tools became widespread, and thus was originally conceptualized as a process to rapidly crowdsource policy solutions from human experts. As shown in the use cases described in this section, combining this traditional crowdsourcing approach with an AI-driven innovation process (mediated by Policy Synth) results in a problem-solving approach that is both more scalable and more effective. We describe the traditional approach in this section, then introduce Policy Synth and explain the value of the augmented workflow that the AI-based toolkit enables.

In 2020, we used the Smarter Crowdsourcing method to advise the Walton Family Foundation and the Bill & Melinda Gates Foundation – philanthropies which fund education projects in the United States – about their grantmaking strategy in the area of non-academic skills assessment in K-12 schools. Without generative AI tools, the process used the following three steps that are common to all of our Smarter Crowdsourcing experiments:

### 3.1 Selecting and Defining the Problem

Based on a literature review and interviews with subject-matter experts, we developed a "Problem Catalog." The Catalog identifies a series of problems that comprise the larger challenge of how to best assess non-academic skills, root causes, and example solutions. The foundation selected a subset of these problems to prioritize. Each problem became the focus of an online convening. Conducting interviews at this stage allows us to gather insights from the collective intelligence of half a dozen (sometimes more) experts, allowing us to understand the "big picture" of the problem at hand and what potential solutions might look like. However, identifying, interviewing and extracting learnings is time-consuming and the diversity of perspectives is limited.

### 3.2 Convening Experts to Identify Solutions

For each problem area selected by the foundation, we curated a list of 80-100 individuals who are experts, including credentialed experts as well as those with lived experience. These experts were identified through an extensive search of relevant literature as well as through recommendations from others working in the field.

For each problem identified by the foundation, we convened a 2-hour, moderated advisory session, using the videoconferencing platform Zoom, where the goal was to identify concrete and specific solutions. Between 25 and 35 experts participated in each convening; the group was curated so as to represent a diverse group of perspectives, including experts in academic curriculum development, psychometrics, career skills pathways, community engagement, and related fields. The conversations surfaced existing solutions as well as ideas for new projects and programs designed to address the problems identified. We ran six convenings, engaging more than 150 experts from six continents. Following each conversation, we spent 5-7 days creating a transcript based on the Zoom recording, writing notes, compiling resources shared, and synthesizing learnings, and then developed findings based on the participants' input.

Engaging such large groups in structured deliberations offers the opportunity to rapidly gather insights, opinions, and new ideas for solutions from experts working on the problems at hand as well as those working in related domains. But there are also many design challenges involved in convening these participatory processes. There is the challenge of ensuring that the group convened has the expertise needed to generate on-topic and high-quality ideas. There is also a need to ensure that the group is diverse and reflects a range of identities, backgrounds and experiences. The conversation must be structured and facilitated in such a way that all participants are encouraged and provided the opportunity to contribute meaningfully to the deliberation. Finally, there are logistical challenges, namely, the large amount of time and labor required to schedule, plan, create the inputs, moderate, and co-create the outputs from these large-group conversations.

### 3.3 Converting Insights into Action

Following the online convenings, the foundation selected a subset of solutions for further exploration. We conducted follow-up surveys, interviews, and research to create a set of recommendations for how to put each solution into action, including costs estimates, budgets, and metrics for evaluation. This step is crucial to ensure that the collective wisdom of the experts is channeled into outputs that are in the form and format that allows the partnering institution to take action upon them. However, converting the insights to action can bring challenges. In past projects, it has taken 3-6 months of research to return final recommendations to the partnering institution. Conducting surveys and interviews and translating the findings from those engagements into detailed memos and reports is simply a laborious and time-consuming process.

Since 2022, some of the more labor-intensive tasks – namely, transcription and notetaking – have been made easier by the availability of out-of-the-box AI assistants. But with the advent of GPT-4, we can now use generative AI to radically reimagine the problem-solving process.

## 4 COMBINING AI AND CI TO SOLVE PROBLEMS COLLABORATIVELY AT SCALE

As shown in the steps above, the typical expert-informed problem-solving process is both time-consuming and labor-intensive. While the Smarter Crowdsourcing approach is designed for efficiency, there is a significant workload involved in conducting research about problems and solutions, curating and convening experts, and synthesizing solutions.

With the release of GPT-4, new possibilities have emerged for leveraging AI and CI together to solve problems more effectively. GPT-4's advanced language understanding, generation capabilities, and ability to process and analyze vast amounts of data have opened up new avenues for integrating AI into collective problem-solving processes. By harnessing the power of GPT-4 and similar AI models, we can enhance the capacity of human groups to generate ideas, evaluate solutions, and make informed decisions.

As described in the following section, Policy Synth helps to solve problems more rapidly by automating many of the research and writing tasks, while providing opportunities for the human guides to provide input throughout the process to ensure the outputs are aligned with the institutions' needs.

## 5 WHAT IS POLICY SYNTH?

Policy Synth is a toolkit designed to automate and scale up the Smarter Crowdsourcing method by seamlessly integrating collective and artificial intelligence. It builds upon the proven success of Smarter Crowdsourcing, which engages the collective intelligence of human experts to solve public problems, by incorporating AI capabilities to enhance the process. Policy Synth leverages artificial intelligence techniques and genetic algorithms to make the Smarter Crowdsourcing method more efficient, effective, and scalable, enabling institutions to develop solutions more rapidly than would be possible with human intelligence alone. The toolkit is built as a JavaScript class-based library, providing a framework for creating AI agent logic flows, application programming interfaces (APIs), and state-of-the-art real-time AI-focused web applications, all designed to streamline and optimize the problem-solving process.

Our goal in creating Policy Synth is to provide institutions access to a set of easy-to-use tools and methods that will increase their ability to solve urgent problems and provide a model process for how institutions can meaningfully engage communities in the problem-solving process.

## 6 SMARTER CROWDSOURCING WITH POLICY SYNTH

The Smarter Crowdsourcing process is designed to channel the collective intelligence of those with expertise about a problem into actionable solutions through curation and convening. In this section, we outline how artificial intelligence can accelerate this process, helping us generate more and better solutions, faster.

As described above, Policy Synth is an AI-based toolkit that accelerates and improves the process of searching for, evolving, ranking, and helping to prioritize among both problems and solutions. Policy Synth augments the stages Smarter Crowdsourcing process as follows:

### 6.1 Selecting and Defining the Problem

Policy Synth accelerates the problem definition phase by automating much of the process of searching for and prioritizing among constituent problems.

The tool is first populated with a "problem statement" – a short paragraph that describes the challenge that the tool is tasked with "solving." The platform will then use AI to automatically: 1) Create a set of queries – questions or phrases that it will plug into a search engine to search for root causes of the problem, 2) Generate a list of problems based on the search of thousands of online sources, and 3) Assign each problem a rating and rank them. The problems are ranked by a set of criteria specified by the user, combined with Elo scoring (See detailed description of the ranking process in the section of this study on "Elo Pairwise Voting."). Once the search and ranking are complete, the tool delivers a list of problems along with the web pages that were scanned in order to identify them.

Artificial intelligence allows us to automatically search a wider range of sources than would be possible through traditional research methods, and to conduct this search more rapidly.

### 6.2 Convening Experts to Identify Solutions

Policy Synth augments the process of identifying solutions, both by aiding in the process of identifying experts to invite to participate in the online convenings, and by automatically developing a separate set of solutions in parallel.

First, Policy Synth can be used to automatically generate a list of solutions in the problem areas of greatest priority. The tool will search the web for solutions that respond to the problems identified previously. After generating hundreds of solutions, Policy Synth will remove any duplicate solutions and filter out the solutions that are not viable. Next, the tool will evolve the solutions using a genetic algorithm. The software combines multiple solutions into one, and then tests how well the new version of the solution fits the problem to see if the improvement should be adopted or rejected. After multiple rounds of crossovers, mutations and ranking, Policy Synth produces a final list of approaches tailored to addressing the problem. (See detailed description of the genetic algorithm in the "Evolutionary Algorithm" section of this study)

This process produces a list of solutions for each identified problem, in an easily readable format with a list of pros and cons for each solution. Each solution is accompanied by a visual illustration from Dalle-3 or other AI image-generation tools.

Second, the toolkit's search functionalities can also accelerate the process of identifying experts to engage in the process of identifying solutions. The list of sources returned by the online searches for problems and solutions – which may include online journal articles, white papers, news reports, and items from other types of online publications – can be useful for identifying those with credentialed expertise and live experience in each problem area, included those who have developed or proposed solutions. For example, the authors of those sources, those cited in reference pages, or mentioned as experts in the text could be included in the list of experts who are invited to contribute solutions. Thus, Policy Synth can help to search for experts more quickly and to cast a wider net compared to manual online searches, yielding more diverse expertise.

### 6.3 Converting Insights into Action

Finally, Policy Synth augments the process of converting ideas into actionable recommendations. Using seed ideas (for example, the solutions identified in the previous step, or solutions identified by the human experts), the toolkit will generate policy proposals based on an online search, then rank and evolve the proposals using a genetic algorithm. This process leverages large-scale web research to search for policy evidence across 22 categories, which is then provided in an easy-to-access format in the user interface. The output is a set of actionable proposals to put the best or highest-priority ideas into practice.

**6.4 Aligning Partner Institutions' Priorities with Communities' Expertise**

In line with the need to tailor the outputs of the problem-solving process to institutions' needs, each stage of Policy Synth is highly customizable.

The instructions used to search for and rank the problems, solutions, or policy proposals are customizable based on the priorities, constraints, and desires of each institutional partner. For example, the instructions could be tailored to prioritize problems that disproportionately impact women or African Americans, to encourage solutions that can be implemented quickly, or to favor policy proposals that can be implemented within the partner institution's jurisdiction.

The outputs produced at each stage can also be manually reviewed and edited. For example, if the system identifies a problem that the partner institution deems outside of its mandate, the problem can be deactivated so that it is not considered when solutions are identified.

While the experiment described in this study focuses on Policy Synth's potential to enhance problem-solving efforts involving groups of experts, the process can be augmented to incorporate input from other groups. The toolkit can be used alongside online engagement tools to combine the collective intelligence of multiple groups. For example, [All Our Ideas](#) – the "Wiki Survey" tool described in the Elo Ranking and Pairwise Voting section of this study – could be used to engage communities in ranking and prioritizing among problems generated during the Selecting and Defining the Problem section of the problem-solving process. Or, the open source [Your Priorities](#) platform could be leveraged to engage communities in a large-scale online dialogue about which solutions to prioritize for implementation. The ability to integrate community input is crucial to ensure that the solutions developed and supported by partnering institutions respond to real and urgent problems that communities experience.

**6.5 Novelty of Approach**

Combining artificial intelligence and collective intelligence in itself is not new; numerous toolkits have been developed which aim to combine CI and AI processes to rapidly develop solutions to public problems [Baeck and Berditchevskaia. 2020].

In existing approaches, AI is largely viewed as a tool to enhance collective intelligence processes, such as dialogues, deliberations, and crowdsourcing efforts. There are several existing platforms which leverage AI to make sense of and draw insights from large volumes of data to gauge communities' opinions on a given topic. These include toolkits designed to aid in analyzing and drawing findings from large-scale online engagements with communities [Rahim, et al. 2024] as well as platforms to synthesize learnings from rich in-person dialogues [Local Voices Network. 2022]. There is also a set of tools and approaches in which AI is used as a tool to facilitate large-scale online deliberations and dialogues [Konya, et al. 2023] [Mendes, et al. 2019] [Moats and Tseng. 2023][Rosenberg, et al. 2023]. In various ways, these approaches endeavor to use AI to improve the outcomes of engagements with groups, whether by enabling engagement with larger groups, improving the efficiency or quality of analysis, or by enabling more coherent or higher quality deliberations and collective dialogues.

While recognizing the value of these methods, Policy Synth takes a different approach. In addition to being a practical tool for policymakers, Policy Synth serves as a demonstration how AI can be used to synthesize the learnings from deliberations (and other types of CI processes) together with findings from research to develop evidence-based policy solutions systematically. Policy Synth is not intended as a replacement for AI tools that improve or scale up engagement with groups; rather, it is a complementary system where outputs from multiple sources, including those from engagements, can be combined with an eye towards developing better solutions.

As described above, the modular design of Policy Synth allows for policymakers to choose at which point in the policymaking process they would like to integrate external input and in what form. For example, a policymaker may be interested in leveraging insights from engagement with experts in one case, while in another case, learnings from an online citizens assembly composed of a representative sample of the population may be of value. As past research has explored, engagement with communities and experts – when effectively designed and executed – can add value at each stage of the policymaking process while also adding legitimacy [Noveck, et al. 2020]

## 7 COMPARATIVE STUDY OF AI-GENERATED AND EXPERT-GENERATED SOLUTIONS

To test how the outputs from the Smarter Crowdsourcing process carried out without generative AI tools compared to those generated using the AI-based Policy Synth toolkit, we conducted a comparative analysis of these two approaches. The goal of this pilot project was to 1) test whether this approach had the potential to truly add value to the Smarter Crowdsourcing process, 2) to identify shortcomings and challenges in the approach, and 3) to lay the groundwork for further, more robust experimentation. This section describes the methodology used and the outputs of this case study.

### 7.1 Comparative Study Design

In 2023, we worked with another large philanthropy to use the Smarter Crowdsourcing method to identify strategies to counter efforts to undermine U.S. elections through sabotage and election denial [Noveck. 2024].

We first created a Problem Catalog outlining 12 problems comprising the large challenge of election violence and denial. We convened online convenings to crowdsource solutions to three of these problems. The philanthropy then selected the two highest priority problem areas for further research. The first area, Misuse of the Legal System, aimed to address the malicious use of litigation and freedom of information (FOI) requests to disrupt election administration and cast doubt upon the validity of election results. The second area, Election-Related Violence, aimed to surface strategies to mitigate the effectiveness of political violence targeting elections. Through online convenings with experts, we developed 14 solutions to the Misuse of the Legal System problem and 13 solutions to the Election-Related Violence problem.

In parallel, we used Policy Synth to search for and generate a list of problems related to election violence and denial. The tool was then seeded with the two priority problems mentioned above and used to generate solutions to the problems of Misuse of the Legal System and Election-Related Violence.

### 7.2 Rating Methodology

To evaluate and compare recommendations, we developed an LLM agent to assess the alignment between solutions generated by human experts and those produced by the Policy Synth process. By examining the similarity between each solution from Policy Synth and the human-generated recommendations, we calculated a percentage similarity for Policy Synth solutions after 15 generations of evolution by the genetic algorithm. This comparison was conducted for each of the seven sub-problems. The analysis and results for two of these problems – Election-Related Violence and Misuse of the Legal System – are described below. (See appendix for links to the results for all seven sub-problems and the rating software is available on GitHub in the Policy Synth repository).

For the Misuse of the Legal System, we compared 65 AI-generated solutions to the list of 14 human-generated recommendations, for a total of 910 comparisons. Similarly, for Election-Related Violence, we compared 65 AI-generated solutions to 13 human recommendations, for a total of 845 comparisons.

*7.2.1 GPT-4 Rating "System" Instructions*

1. You are an expert in analyzing how well a solution matches requirements
2. Compare the key points in each requirement with the key points in the solution
3. If solution does more than required then that is fine
4. Always and only output the following JSON format: [ { requirementTitle, solutionCoversPercent } ]

*7.2.2 Example input to the LLM rating agent from one comparison:*

**Human "Requirement"/Recommendation**: "Invest in civic engagement projects that get citizens more involved in elections, helping to educate and build trust in the electoral process."

**Evolved AI Solution initially sourced through automated web research:** "Philanthropic organizations should invest in initiatives that build trust in the electoral process. This includes supporting voter education campaigns that emphasize the importance of peaceful elections, the measures in place to ensure election integrity, and the consequences of election-related violence. Funding research to understand the factors that undermine trust in elections and developing strategies to address these issues is also crucial."

**GPT-4 Rating:** 100% (In other words, the AI solution includes 100% of the core idea of the human recommendation. This is meant to evaluate the meaning or sentiment behind the recommendation, not whether the word choice or phrasing is an exact match.)

*7.2.3 Example Ratings from Election-Related Violence topic*

**Problem statement:** "Election-Related Violence: A free, fair, and vibrant democracy depends on the ability to hold peaceful elections, including running for office, casting, counting and certifying votes free of violence and intimidation. Yet, during recent election cycles, extremist groups and individuals have used violent acts and threats to intimidate political opponents, disrupt electoral processes, and cast doubt on the fairness and legitimacy of election results."

Table 1: Example of three pairs of comparisons between AI-generated solutions and crowdsourced recommendations for Election-Related Violence topic.

| Human-Generated Recommendation | AI-Generated Solutions from the Web | GPT-4 rating (% of the human recommendation included in AI solution) |
|---|---|---|
| Invest in civic engagement projects that get citizens more involved in elections, helping to educate and build trust in the electoral process. | Philanthropic organizations should invest in initiatives that build trust in the electoral process. This includes supporting voter education campaigns that emphasize the importance of peaceful elections, the measures in place to ensure election integrity, and the consequences of election-related violence. Funding research to understand the factors that undermine trust in elections and developing strategies to address these issues is also crucial. | 100% |
| | US philanthropic organizations can invest in the development and deployment of a comprehensive digital platform that facilitates real-time reporting of election-related violence and threats. This platform can also provide verified election-related information and resources, thus serving as a countermeasure against misinformation. Citizens, election officials, and law enforcement agencies can use this platform to report incidents, access accurate information, and coordinate responses, ultimately promoting a safer and more informed electoral environment. | 80% |
| | US philanthropic organizations should fund a comprehensive anti-violence training program for election stakeholders, including community-based organizations, civil society organizations, and political parties. The training should focus on identifying potential threats and developing robust security measures for election periods. In conjunction, a mobile alert system should be established to facilitate immediate reporting of incidents and threats of election-related violence. Through the integration of these two components, stakeholders can effectively prevent and manage election-related violence while enabling citizens and election officials to report and respond rapidly to instances of violence or intimidation, fostering a safer electoral environment. | 50% |

*7.2.4 Example Ratings from Misuse of Legal System topic*

**Problem statement:** "Misuse of the Legal System: In recent elections, saboteurs have misused the administrative and legal systems in a coordinated effort to obstruct election administration and sow doubt in the outcomes of fair elections in the minds of the public."

Table 2: Example of three pairs of comparisons between AI-generated solutions and crowdsourced recommendations for Misuse of the Legal System topic.

| Human Generated Recommendation | AI Generated Solutions from the Web | GPT-4 rating (% of the human recommendation included in AI solution) |
|---|---|---|
| Encourage organizations and bodies with disciplinary authority to take swift action against groups, attorneys, and law firms who have advanced malicious litigation. | Philanthropic organizations should champion legal reforms aimed at shielding election administrators from groundless lawsuits and excessive FOIA requests. In collaboration with civil society organizations, community-based organizations, and legal advocacy groups, they should lobby for laws that penalize those misusing the legal system to hinder election administration. These reforms should also facilitate the swift dismissal of baseless lawsuits and FOIA requests. | 100% |
| | Philanthropic organizations should fund legal advocacy groups to monitor and counteract frivolous FOIA requests and lawsuits aimed at disrupting election administration. These groups could also offer legal support to election administrators, assisting them in navigating these challenges and maintaining the election process's integrity. | 70% |
| | Philanthropic organizations should support legal advocacy groups in defending against unnecessary FOIA requests and lawsuits that obstruct election administration. They can do this by providing resources for legal defense, advocating for legislative changes to discourage frivolous actions, and raising public awareness about the misuse of the legal system in elections. This action directly tackles the problem of legal system misuse in elections. | 70% |

## 7.3 Results and Discussion

The comparative analysis showed that there was a high level of similarity between the Policy Synth solutions, sourced via large scale automated web research and the crowdsourced expert solutions. Among the 14 crowdsourced recommendations for the Misuse of the Legal System topic, Policy Synth generated at least one solution that was 100% match for eight of the recommendations. Similarly, Policy Synth yielded at least one solution with a 100% match for 10 out of the 13 recommendations in the Election-Related Violence topic. As shown in the tables below, Policy Synth produced 132 solutions for the Misuse of the Legal System problem area and 64 solutions for the Election-Related Violence problem area that the LLM judged to be a 100% match with at least one of the human-generated recommendations. In other words, the toolkit produced many high-quality, on-target solutions that mirror those generated by human experts.

Table 3: Comparison of human-generated recommendations with AI-generated solutions for Misuse of the Legal System problem area.

| Human-Generated Recommendations | Number of AI-Generated Solutions Scored 100% by GPT4 | Number of AI-Generated Solutions that include >50% & <100% of the core human recommendation |
|---|---|---|
| Recommendation 1: Encourage organizations and bodies with disciplinary authority to take swift action against groups, attorneys, and law firms who have advanced malicious litigation. | 2 | 7 |
| Recommendation 2: Invest in legal education to better teach professional responsibility and support professional development opportunities for students in order to train the next generation of lawyers to fulfill their role as guardians of democracy, elections and the Constitution. | 0 | 6 |
| Recommendation 3: Invest in capacity building, and legal advocacy to better protect election offices from malicious litigation. | 45 | 1 |
| Recommendation 4: Invest in resources to support groups who track, report on, or carry out legal advocacy challenging malicious litigation. | 26 | 13 |
| Recommendation 5: Invest in research to understand how the courts can more effectively deal with malicious litigations. | 0 | 4 |
| Recommendation 6: Engage key stakeholders in the process of developing solutions to the problem of malicious litigation. | 23 | 11 |
| Recommendation 7: Invest in policy research to understand how potential changes to FOI laws and policies will impact public access to information, and in particular, the negative knock-on consequences of policy changes. | 0 | 0 |
| Recommendation 8: Invest in training, data infrastructure, staffing, and other forms of capacity building that enable election offices to proactively release more election-related information, which can allow offices to head-off vexatious or duplicative requests. | 0 | 3 |
| Recommendation 9: Invest in training, data infrastructure, staffing, and other forms of capacity building that enable election offices to process and respond to FOI requests more efficiently and effectively. | 0 | 1 |
| Recommendation 10: Invest in research and policy advocacy to pilot administrative changes designed to limit vexatious FOI requests. | 6 | 3 |

| Human-Generated Recommendations | Number of AI-Generated Solutions Scored 100% by GPT4 | Number of AI-Generated Solutions that include >50% & <100% of the core human recommendation |
|---|---|---|
| Recommendation 11: Fund research and further engagement with experts to fully understand the scope of the problem. | 1 | 0 |
| Recommendation 12: Invest in improving administrative processes, filling funding gaps, and providing election offices with the resources they need to continue to run elections with minimal errors and disruptions. | 0 | 14 |
| Recommendation 13: Invest in research, and provide sustained funding to on-the-ground organizations, to enable the election protection groups to anticipate, and prepare to respond, to future election subversion campaigns. | 11 | 19 |
| Recommendation 14: Engage communities and build coalitions to ensure that the voices of key stakeholders are represented in funding decisions related to election administration. | 18 | 12 |
| **Total** | **132** | **94** |

Table 4: Comparison of human-generated recommendations with AI-generated solutions for Election-Related Violence problem area.

| Human-Generated Recommendations | Number of AI-Generated Solutions that include 100% of the core human recommendation | Number of AI-Generated Solutions that include >50% & <100% of the core human recommendation |
|---|---|---|
| Recommendation 1: Invest in Early Warning Response Systems (EWRS) – tools which use data and research findings to predict instances of violence – with an emphasis on rapidly delivering intelligence about threats to elections groups working at the local level. | 5 | 15 |
| Recommendation 2: Invest in de-escalation training for election officials to equip them with strategies to prepare for and effectively respond to potentially violent incidents. | 5 | 26 |
| Recommendation 3: Explore how changes to law – such as better protections for the personal identifying information of election officials and judges, or introducing stiffer penalties for threats targeting public servants – could prevent acts of violence | 0 | 13 |

| Recommendation | | |
|---|---|---|
| Recommendation 4: Invest in building stronger relationships among elected officials and law enforcement to allow law enforcement to better anticipate, prepare for, and respond to violent threats and actions. | 6 | 13 |
| Recommendation 5: Invest in longitudinal data collection about violent threats, attitudes, political beliefs that justify violence, and other data related to violence to better understand the risk of violence and make better predictions. Collecting data about violent acts alone is insufficient. | 0 | 20 |
| Recommendation 6: Fund a multidisciplinary research collaborative to develop and test effective messaging strategies, borrowing from successful campaigns in other domains, and designed to combat the spread of anti-democratic ideas. | 0 | 30 |
| Recommendation 7: Fund efforts to proactively respond to and counter online hate speech that fuels violence. | 2 | 18 |
| Recommendation 8: Support more transparent, resilient, and bipartisan election administration practices that build public confidence in electoral procedures and outcomes. | 2 | 24 |
| Recommendation 9: Invest in civic engagement projects that get citizens more involved in elections, helping to educate and build trust in the electoral process. | 22 | 21 |
| Recommendation 10: Explore how to engage with social media companies in developing and implementing solutions to stop the spread of misinformation on their platforms which fuels election violence. | 1 | 23 |
| Recommendation 11: Advocate for Congress to take a leading role in combating violent threats and acts against election officials. | 1 | 24 |
| Recommendation 12: Center local communities – in particular, communities of color and other groups who have historically been targeted by political violence – in grantmaking decisions around election violence | 10 | 21 |
| Recommendation 13: Provide sustained funding that allows groups to work on election violence issues continuously across election cycles. | 10 | 29 |
| **Total** | 64 | 277 |

The toolkit also produced new solutions that were not identified by the human experts. For example, to address the Misuse of the Legal System problem, for example, Policy Synth identified new recommendations such as establishing a legal defense fund for administrative officials and mental health support for election workers. For the Election-Related Violence problem area, Policy Synth suggested novel solutions such as integrating geotagging into existing mobile alert systems by integrating a geo-tagging feature to allow threats to be pinpointed more rapidly.

The results of the experiment shows that Policy Synth can be an effective tool both for sourcing existing solutions, and for generating novel ideas.

## 7.4 Limitations and Future Research

It should be noted that the comparative analysis was not intended as a robust, scientific experiment. Rather, it was conceived as a pilot test case with the goal of informing future development work and higher-quality experimentation.

One key limitation is that, due to time and capacity constraints, the comparative analysis relied on LLMs to evaluate the similarity of the expert and AI-generated solutions. While there is some debate over to what extent LLMs are capable of semantic understanding, it is clear that significant shortcomings remain when it comes to LLMs' ability to understand logic and meaning in context, as is required to engage in complex comparisons [Wu, et al. 2024] [Yan, et al. 2024].

Further, basing the evaluation primarily on the number of similar statements does not capture the whole picture. As described above, there may be value in both statements that are similar to those generated by experts (as validation for the AI approach) and those that are dissimilar (for assessing the novelty or added value of the AI approach). This nuance may not be well-captured by our evaluation metrics.

We plan to conduct a future experiment, the design of which will address these limitations. The experiment will leverage human reviewers, in addition to LLMs, to evaluate the solutions.

We will also compare the similarity between the LLM evaluations of the solutions and the human evaluations to better understand how well LLMs can provide this role at scale. We will also develop more robust evaluation metrics that will be used to score the machine-generated ideas rather than simply comparing them to human-generated solutions while also evaluating the quality of the AI system.

## 8 TECHNICAL DESCRIPTION OF POLICY SYNTH

In this section, we outline the design considerations that went into the platform's architecture. We also explain some of Policy Synth's key technical components – including AI agents, Elo scoring, pairwise voting, and genetic algorithms – and how they function within the platform.

### 8.1 Inspiration for Technical Design

The technical design of the Policy Synth toolkit was inspired by our 35+ years of experience in building AI systems, including very early chatbots and high-profile AI driven video games, and insights from many directions.

One source of technical inspiration was the paper "Language Models are Weak Learners," [Manikandan, et al. 2023] which highlighted the imperfections of AI models that continue to make errors, as AI models have done from the start. However, there have always been strategies to mitigate the fact that AI models are "weak learners." This insight has been crucial, steering us towards strategies that enhance the performance of current generation AI tools. These strategies include always simplifying tasks for large language models (LLMs) as much as possible, breaking complex "thoughts" into simpler "thoughts", employing pairwise ranking for efficient selection, and incorporating review loops for iterative refinement.

Another inspiration was a thought-provoking piece in *The Economist*, "It's not just a fiscal fiasco: graying economies also innovate less," which underlined the importance of "fluid intelligence" – a concept in psychology which describes "the general ability to reason, to flexibly engage with the world, to recognize patterns, and to solve problems in a manner that does not depend upon specific previous knowledge or experience" [Cochrane, et al. 2019] in fostering innovation. This led us to evolutionary algorithms. Such algorithms, by their nature, could replicate aspects of human creative thinking.

Together, these insights shaped Policy Synth is intended as a tool not just for scaling problem-solving efforts but also for enhancing the creative capacity of collective intelligence through AI. It is an attempt to blend the precision of algorithms with the nuanced understanding of human experts, aiming for a balance between technological efficiency and the rich, diverse insights that come from collective human engagement.

### 8.2 About AI Agents

AI agents are a key component of Policy Synth's approach to System 2 thinking (described below). In the context of computational systems, AI agents are autonomous or semi-autonomous entities that interact with digital environments to achieve specific goals or tasks. These agents can range from simple rule-based algorithms to complex systems like Large Language Models (LLMs), capable of processing and generating text and performing simple reasoning. The essence of AI agents lies in their ability to perceive their environment through data, make decisions based on this data, and act upon these decisions to fulfill predefined objectives. In Policy Synth AI agents are used extensively for different tasks.

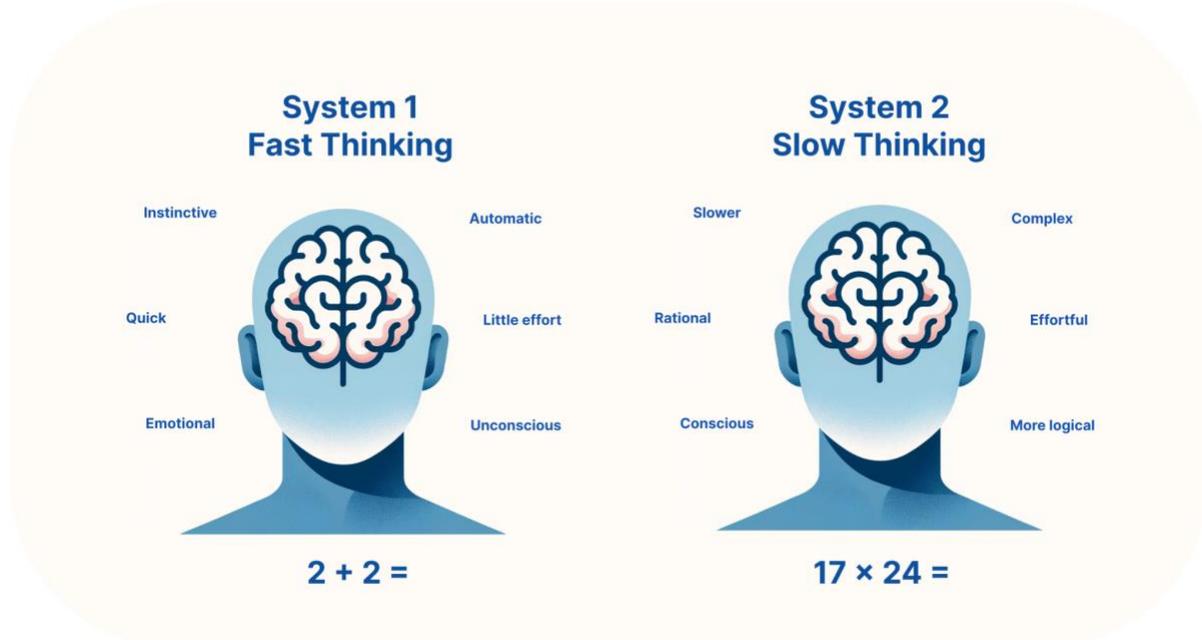

Figure 1: Illustration comparing System 1 vs System 2, types of cognitive processing.

### 8.3 Fast and Slow Thinking

Fast (System 1) and Slow (System 2) thinking, terms popularized by psychologist Daniel Kahneman, offer a useful framework for understanding how AI, particularly LLMs like GPT-4, operates [Kahneman. 2013]. LLMs are often likened to System 1 thinking [Hagendorff, et al. 2023] where cognition occurs almost instantaneously, without deliberate, conscious effort, much like a human's immediate response to simple arithmetic such as "what is 2+2?" While the best LLMs are also capable of limited form of System 2 thinking then this analogy suggests that LLMs excel in producing responses quickly, leveraging vast databases of information to generate answers that seem instinctive, much as System 1 thinking does for humans in certain contexts.

System 2 thinking represents a more deliberate, analytical form of reasoning that humans employ for more complex tasks, such as solving "what is 17*24?" without the aid of calculators. This level of cognitive processing requires significant effort, attention, and mental manipulation of information.

The aspiration to develop AI agents capable of fully autonomous System 2 thinking – a form of AI that can engage in deep methodical problem-solving – remains a work in progress. While advancements are being made towards creating AI that can mimic this type of cognition [Yao, et al. September, 2023], achieving fully autonomous System 2 thinking in AI is an ongoing challenge, underscoring the complexity of human intelligence.

In contrast to striving for fully autonomous System 2 capabilities with AI, Policy Synth approaches the challenge by manually constructing and integrating System 2 thinking processes through an AI agent framework. This methodological choice allows for the deliberate orchestration of AI processes that implement a deeper System 2 thinking by combining specialized agents in a structured manner.

**8.4 Multi-Scale Policy Synth Agents**

Policy Synth employs an innovative approach through the integration of multi-scale System 2 AI agent thinking processes. This approach is grounded in the utilization of numerous straightforward System 1 LLM agents/prompts, which are carefully organized into workflows that operate at various scales.

Within this framework, the Text Compression Agent serves as a straightforward example of an agent deployed (see diagram below). This agent takes text in and outputs the same text with as few words as possible while still keeping all detail, nuance and tone. This agent is supported by three parallel Validation Agents designed to ensure the output's Correctness, Completeness, and to detect Hallucinations. These agents operate within a feedback loop, critically evaluating and refining the compression prompt's output to correct errors, which are present in approximately 10% of the main Text Compression Agents outputs in this specific and complex task.

The integration of these validation processes enhances the reliability and accuracy of the agent's output, illustrating a key step in the process of data refinement. Furthermore, the Text Compression Agent is not isolated but is part of a larger system, where it can contribute to more complex agents focused on tasks such as data ingestion or custom data cleanup workflows.

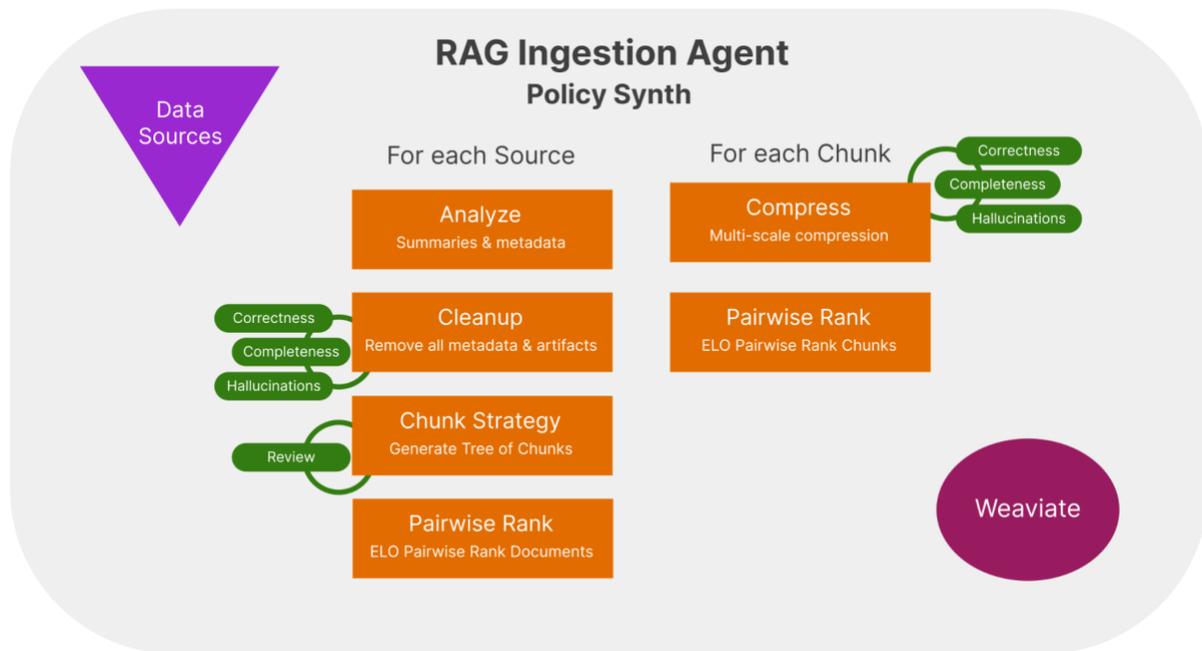

Figure 2: Example of a simple Policy Synth data Ingestion Agent using System 1 thinking LLM prompts in a System 2 thinking process at multiple scales. Takes in unstructured data sources in multiple formats and outputs them clean and compressed into a vector database.

### 8.5 Large Scale Automated Web Research

Policy Synth enhances the policymaking process through AI-driven large-scale automated web research with Google or Bing search, then scanning thousands of websites with LLMs. Root causes discovery is a key example application. Those automated web searchers and scanning are versatile enough to explore any topic, encompassing various types of content and engaging multiple AI agents exploring separately highly detailed aspects of the data.

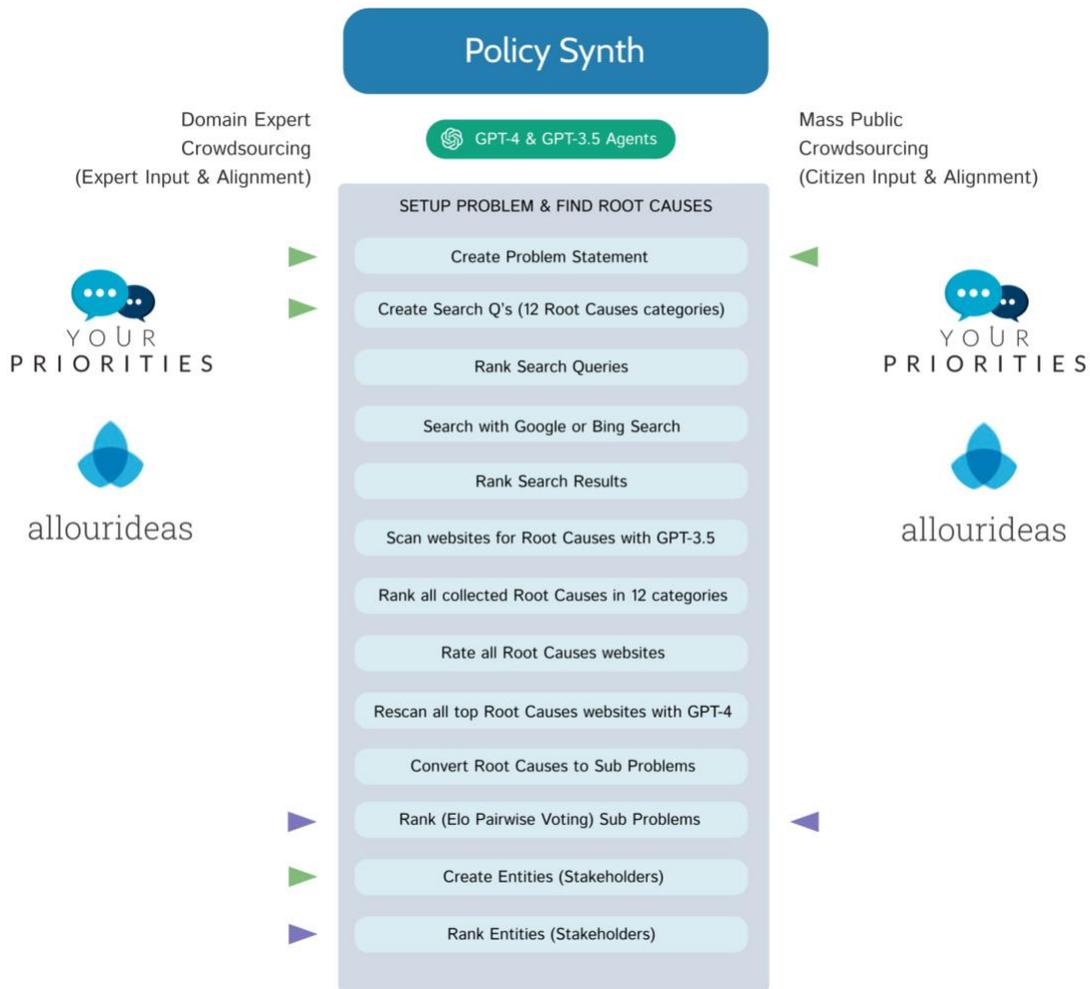

Figure 3: Example of large-scale automated web research AI agent workflow for root cause discovery, with options for direct inputs from experts and citizens at different stages through Your Priorities for Idea Generation and All Our Ideas for "human" pairwise ranking of problems and ideas.

The provided diagram outlines a systematic approach, beginning with the creation of a problem statement, which can be adapted to any subject. From this starting point, Policy Synth automatically generates a multitude of search queries. These are not limited to uncovering root causes but can be tailored to specific needs, whether they are scientific inquiries, data analysis, or current events. The AI then ranks these queries to optimize the search process.

Upon executing the search through standard engines like Google or Bing, the AI ranks the results, directing the fast and inexpensive GPT-3.5 agents to scan for pertinent information, which is subsequently organized into relevant categories. The more capable GPT-4 agents scan high-priority sources to refine and enhance the gathered data.

This process translates into identifying and ranking sub-problems or various other types of information, which are then evaluated using an Elo pairwise voting system to enable us to use the best LLM outputs only. The final step in this phase is to identify and rank stakeholders or other relevant entities, thereby converting a broad spectrum of internet data into a curated, actionable format.

**8.6 Elo Scoring and Pairwise Voting**

To rate and rank-order problems and solutions, Policy Synth uses a process known as "Elo ranking."

Originally developed for assessing the relative skills of chess players, Elo ranking is an algorithmic approach to rate and rank a list of options or candidates. In this method, options or candidates are treated as competitors in a series of pairwise contests, where voters express a preference between each pair. The Elo algorithm then adjusts the ratings, or "scores," of these options based on the outcomes of these contests, taking into account the relative ratings of the options at the time of the vote. The magnitude of rating adjustment depends on the expected versus actual outcome, allowing the system to dynamically rank options in an efficient way.

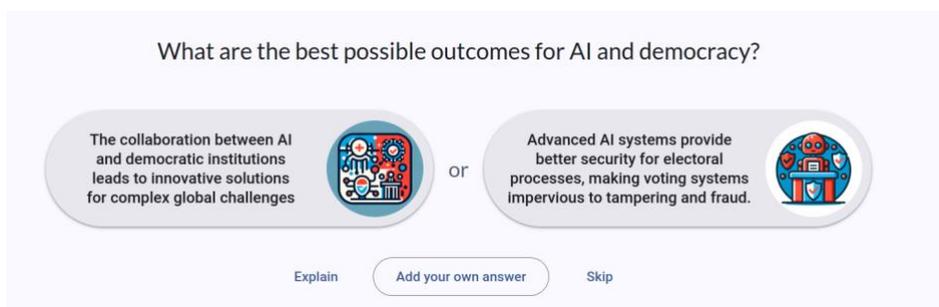

Figure 4: Example of human pairwise voting using the All Our Ideas platform. You can find an example of a AI Pairwise Voting Agent/Prompt in the Appendix

In the context of Policy Synth, Elo-based pairwise voting plays a critical role in enhancing the quality of outputs generated by LLMs, and we use it to provide the fitness function for the genetic algorithm.

Just as humans find it easier to compare two items at a time rather than multiple items, LLMs exhibit better performance with simpler ranking tasks [Qin, et al. 2023]. For a human sorting a small list, such as 10 items, the task might be managed intuitively. However, when faced with a larger list, say of 100 items, a more systematic approach is required, which is akin to engaging System 2 thinking. Similarly, LLMs are more effective when employing a straightforward method of comparison, like pairwise ranking, which minimizes complexity and focuses the reasoning process on one comparison at a time.

By adapting Elo-based pairwise voting, Policy Synth utilizes it, for example, to compare two search queries or to compare pairs of potential solutions. Voters – either human participants or AI – indicate their preference between each pair. Solutions are treated as competitors in these matchups, with the algorithm assigning higher scores to those preferred by the voters, effectively simulating a tournament of ideas where the strongest candidates emerge based on collective assessment.

This dynamic ranking mechanism serves multiple purposes including ranking solutions. Firstly, it helps prioritize solutions that are more likely to be effective, as determined by the collective intelligence of the participants. Secondly, it serves as an integral part of the genetic algorithm's fitness function within Policy Synth. Solutions that score higher in pairwise voting have a higher "fitness" and are therefore more likely to be retained and combined in subsequent generations of solution development.

Pairwise voting has proven its utility beyond the scope of Policy Synth, finding applications in areas ranging from optimizing decision-making processes in group settings to enhancing machine learning models. Pairwise voting informs the design of All Our Ideas, an online engagement platform that is used by institutions to facilitate large-scale dialogues with the public. Since launching in 2010, [All Our Ideas](#) has been used more than 27,000 times to facilitate dialogues on topics ranging from improving road safety in Brazil to understanding Americans' concerns about the use of "big health" data [Gambrell 2020]. Pairwise voting has also been the subject of academic research. For instance, Limayem and Yannou [2007] examined judgmental inconsistencies in pairwise comparisons within groups, underscoring the method's significance in deriving more accurate collective decisions. Similarly, [Zucker. 2020] explored social choice mechanisms for resolving incomplete and intransitive pairwise votes, contributing to the development of more robust decision-making frameworks. Moreover, the work of [Mencía and Fürnkranz. 2008] in efficient voting prediction for pairwise multilabel classification highlights the approach's adaptability to complex machine learning challenges, offering a pathway to reduce computational demands while maintaining predictive accuracy.

**8.7 Evolutionary Algorithm (Genetic Algorithm)**

Evolutionary algorithms, including genetic algorithms, are a family of algorithms inspired by the principles of natural selection and genetics, used to solve optimization and search problems within computer science. These algorithms work by mimicking the process of natural evolution, employing mechanisms such as selection, mutation, and crossover (recombination) to evolve a population of candidate solutions towards an optimal solution over generations. Initially, a diverse population of possible solutions is generated. Through the iterative process of selecting the fittest individuals based on a defined fitness function and applying genetic operators like crossover and mutation to produce offspring, the algorithm seeks to improve the population's overall quality. Over successive generations, the population evolves, and ideally, a solution emerges that is close to the optimum for the problem at hand.

In Policy Synth, the genetic algorithm is a versatile tool, utilized for various aspects of policy development. The evolutionary aspect of Policy Synth is manifested in its ability to not just find ideas but also to develop them further. By applying an Elo-based pairwise voting system, the most effective solutions are identified and carried forward. Ideas that are not viable are pruned, while promising ones are enhanced through pros and cons analysis, further refined through successive generations of mutation and crossover within the algorithm.

Incorporating human judgment is also a key aspect, allowing the genetic algorithm to tailor its fitness function to human preferences and societal requirements. Human input can guide the algorithm, ensuring that the evolved policies not only are optimized according to only quantitative data but also fit with qualitative human values and practicalities.

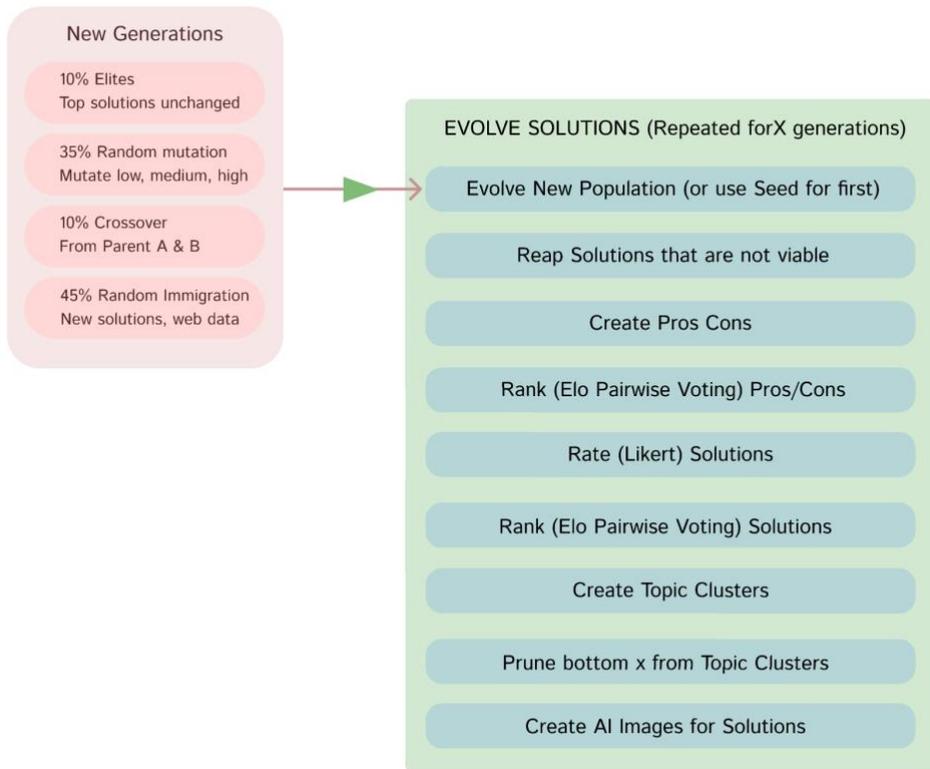

Figure 5: Example of AI agents flow-chart for evolving solutions using pairwise voting for fitness function.

Incorporating human-sourced intelligence, artificial intelligence, and the procedural intelligence of evolutionary algorithms enables a multidimensional fusion of cognitive processes, each operating at varying scales and complexities. Human-sourced intelligence brings to the table the nuanced understanding, ethical considerations, and creative problem-solving capabilities intrinsic to human cognition. Artificial intelligence offers computational efficiency, pattern recognition, and data processing at scales and speeds unattainable by humans. Meanwhile, the intelligence of evolutionary algorithms lies in their capacity to navigate vast solution spaces through iterative optimization, implementing the trial-and-error process of natural evolution.

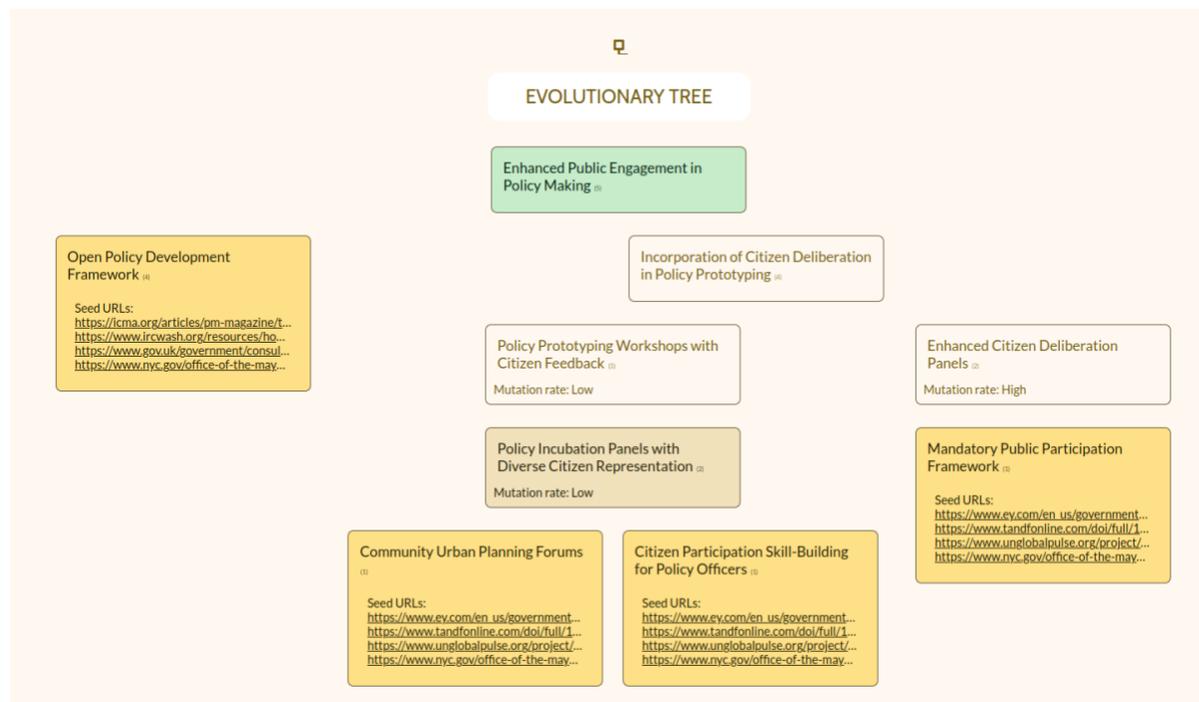

Figure 6: Example solution evolved over 5 generations from human solutions found via automated web research.

Our practical research furthers the understanding of the potential integration between genetic algorithms and large language models, focusing on the application and prospective advancements within this interdisciplinary field. The convergence of computational algorithms and LLMs is supported by a range of studies, as indicated by computational research references [Romera-Paredes, et al. 2024] [Guo, et al. 2023] [Hemberg, et al. 2024].

## 9 ETHICS

Policy Synth embodies a principled approach to AI development, prioritizing human-centric design, collaboration and inclusivity, open-source and nonprofit principles. We elaborate on each of the ethical considerations that guide the project's development below.

### 9.1 Human-Centric AI

Policy Synth is designed with a human-centric approach, recognizing AI as a tool to augment human intelligence and creativity rather than replace it. This perspective ensures that the technology is developed and applied in ways that enhance human decision-making processes, respect human dignity & agency, and foster a symbiotic relationship between humans and AI. Such an approach is critical to addressing ethical concerns related to autonomy, accountability, and the potential dehumanization that can arise from the misuse of AI technologies.

### 9.2 Collaborative and Inclusive Development

Policy Synth's development process is intentionally inclusive, involving stakeholders from various fields to ensure a diverse range of perspectives and expertise. This collaborative approach not only enriches the tool's development but also aligns with ethical principles of inclusivity and equity.

### 9.3 Open Source Foundation

Policy Synth is developed under an open-source model, consistent with the ethical stance of ensuring transparency and collaboration in technological development. This model supports the broader scientific and technological community's access to, and improvement of, our tools. By releasing Policy Synth with an [MIT license](#) – which permits others to reuse, adapt, and republish the code without restrictions – we underscore our dedication to offering a platform that can be audited, modified, and implemented by others.

### 9.4 Non-Profit Civic Innovation

The development and dissemination of Policy Synth are driven by a non-profit mission, aimed at maximizing public good rather than private gain. This approach is crucial in maintaining the tool's integrity and ensuring it serves the broader objectives of enhancing public problem-solving and policy development. By prioritizing societal benefits over commercial interests, Policy Synth adheres to ethical principles that prioritize societal impact and the democratization of AI technologies.

## 10 LIMITATIONS OF THE EXISTING TOOL AND PLANS FOR FUTURE DEVELOPMENT

### 10.1 Current Limitations

One of the primary challenges facing Policy Synth is the considerable expense associated with using advanced AI models like GPT-4 for extensive web research and the genetic evolution of large populations across numerous subproblems. The costs for such operations can range significantly, from a few thousand to tens of thousands of dollars. However, it's worth noting that since the start of the Policy Synth project in June 2023, there has been a substantial reduction in these expenses, approximately 75%. Continuous efforts are made to optimize the code to ensure minimal expenditure without compromising the quality of outcomes and the cost of advanced AI models keeps on going down.

### 10.2 Future Directions: Research and Impact

As AI development continues to accelerate, it stands as one of the most significant opportunities for society to address the challenges of information overload and the often-paralyzing uncertainty about the future. By effectively managing and making sense of the deluge of data available, AI can help policymakers navigate complex societal challenges with greater confidence. Moving forward, the promise of AI-enhanced governance extends beyond improved efficiency and inclusivity in decision-making. It harbors the potential for more stable politics, where informed and participatory governance can reduce societal anxiety about the future. Through the judicious integration of AI in the policy-making process, we can aspire to a future characterized by not only more effective government action but also by a more engaged and reassured public, confident in the resilience and adaptability of their political systems to meet emerging challenges head-on.

Another consideration is how to account for invisible biases. These may include the risk of replicating biases in existing literature, and additional biases introduced by LLM agents in their rating, prioritization, and summarization of research. Future experiments will explore how potential biases in our approach can be better identified and accounted for.

In future projects, we plan to fully combine the AI and CI processes, rather than running them in parallel. Insights from large-scale engagements with communities will be used to inform the identification and prioritization of the problems that the Policy Synth toolkit is tasked with solving. Future experiments should explore the value of Policy Synth and other models for incorporating the expertise of communities into the process of identifying problems, generating solutions, implementing solutions, and evaluating the impact.

Finally, further experimentation is needed to also explore how to overcome the challenges and limitations of the problem solving approaches described in this study. One key challenge is the limited capacity of institutions to organize and deploy time-intensive and labor-heavy processes like Smarter Crowdsourcing – with or without automation. As generative AI tools continue to be deployed rapidly across various aspects of society, there is a risk that the technological capacity gap between governments and the private sector will grow. Research should explore whether and how Policy Synth and other AI tools can be used to narrow, rather than continuing to widen, this gap.

**10.3 Future Directions: Technical**

Looking forward, the Policy Synth project plans to undertake significant further developments. Already released in 2024, we have devised a new framework for a chatbot utilizing Policy Synth agents. This framework aims to establish automated data ingestion pipelines, aligning with the foundational principles of Policy Synth to enhance process efficiency and broaden the project's scope. Policy Synth Engineer, an automatic programming solution for Typescript projects, has entered the early stages of development, already completing GitHub tickets fully autonomously, and is estimated to help scale up future development pace of Policy Synth considerably.

Incorporating insights from biology, particularly Michael Levin's work [Lagasse and Levin. 2023] [Watson and Levin. 2023], offers a nuanced perspective on computational models. Levin's and others research into bioelectric phenomena and multi-scale collective intelligence parallels some of the adaptability of biological systems with the structured yet flexible framework of Policy Synth's multiscale agents and genetic algorithms. While both systems exhibit a level of adaptability and complexity, a key difference emerges in the goal-driven behavior of biological agents compared to the more task-oriented focus of Policy Synth agents. This distinction prompts a research question: Can some ideas related to the goal-driven principles observed in multi-scale biological intelligence be integrated into Policy Synth's computational LLM agents to enhance their functionality?

**11 CONCLUSION**

In the progression from our institutional experience with collective intelligence projects that did not have the input and assistance of AI, to our work on Policy Synth, one clear and obvious outcome of integrating AI is increased efficiency. AI helped us to summarize and extract the learnings from two hours of simultaneous talking and typing in minutes, rather than days. AI enabled research of inputs on a topic and assembling of source documents for the human agents to consider almost instantly, reducing the expense of these "back office" research work that these sorts of projects would incur if they did not have AI assistance.

The integration of humans and AI in decision-making processes represents a promising path for enhancing governmental decision-making. By harnessing the capabilities of AI through platforms like Policy Synth, there is a unique opportunity to process vast amounts of data, enabling more informed, equitable, and participatory decision-making. The collaborative synergy between human intuition and AI's analytical prowess can lead to increasingly sophisticated insights into the most effective use of public resources. This partnership aims to cultivate a more stable political climate where decisions are made with a comprehensive understanding of societal needs, ensuring that no group is left behind.

Beyond efficiency, there is compelling evidence from our research and in recent work published in *Nature* [Hubert, et al. 2024] that in creative attempts to conceive of divergent elements in problem-solving, human and AI agents produce complementary results. Our results show that AI can generate responses that mirror those produced by human experts, as well as novel solutions that introduce new ideas, adding value to the problem-solving process. This outcome underscores the value of bringing together human collective intelligence and generative artificial intelligence, in producing the most comprehensive list of innovative and actionable solutions to policy problems.

## A   APPENDICES

**EXAMPLE OF AUTOMATED POLICY SYNTH TESTS**

While Policy Synth is designed for human experts and public input at various stages, we have run tests with a fully automated Policy Synth agent chain over periods of 5 to 20 days to identify problems, search for human solutions, evolve ideas and do automated policy evidence web research. This was without any human intervention aside from writing the problem statements. All solutions generated through this automated AI agent process still all come from human sources through extensive large scale automated Google searches and website scanning of thousands of websites.

**Example 1: Democracy in Distress**

Link to live test results**:** https://policy-synth.ai/projects/1/

**Problem Statement:** Liberal democracies are grappling with an overarching issue of declining effectiveness and stability, which is leading to widespread citizen dissatisfaction and threatening the fundamental principles of democratic governance.

Figure A1: Top Sub Problems Identified:

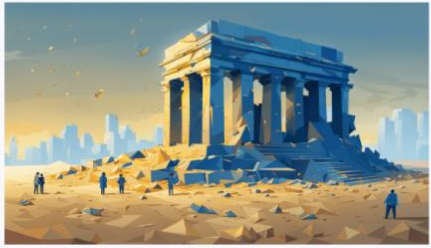

Liberal democracies are grappling with an overarching issue of declining effectiveness and stability, which is leading to widespread citizen dissatisfaction and threatening the fundamental principles of democratic governance.

Sub Problems and Solutions

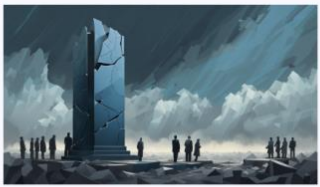

**Erosion of Public Trust**

There is a palpable erosion of citizens' trust in democratic processes and institutions, posing a risk to the credibility of liberal democracies.

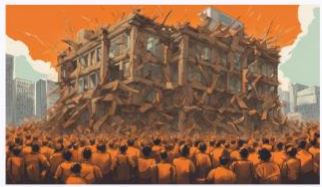

**Ineffectiveness of Democratic Institutions**

Democratic institutions are consistently failing to deliver on policy expectations, leading to public disapproval.

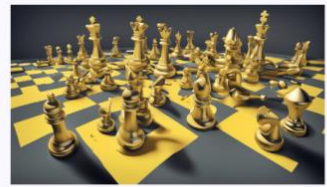

**Persistent Political Instability**

Consistent instability due to frequent changes in government or policy introduces a climate of uncertainty and inconsistency.

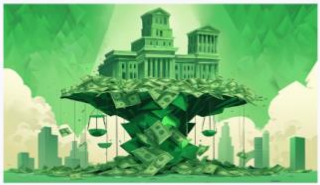

**Prevalence of Corruption**

Widespread corruption and misuse of power erode the effectiveness and credibility of democratic institutions.

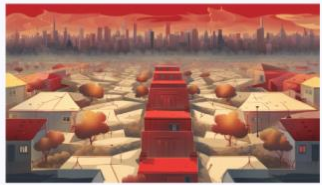

**Rise in Political Polarization**

The surge in political polarization is causing societal divisions and fostering conflict among different groups within the society.

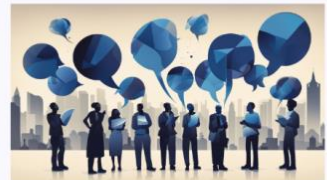

**Spread of Disinformation**

An unchecked spread of disinformation is impairing citizens' ability to make informed political choices and engage in rational discourse.

Figure A2: Example search queries generated by Policy Synth

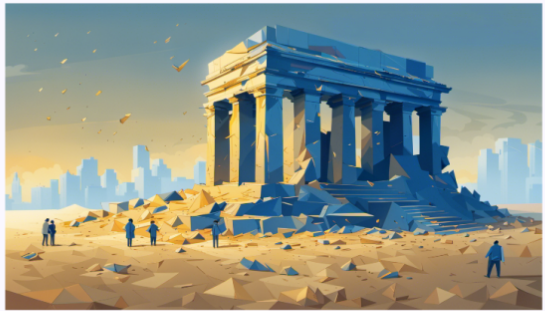

Figure A3: Example of websites scanned.

**Webpages scanned for solutions to problem statement**

general

The populist challenge to liberal democracy
https://www.brookings.edu/articles/the-populist-challenge-to-liberal-democracy/

The Key to Revitalizing Liberal Democracy: Think of It As ...
https://www.beyondintractability.org/ccg-democracy-is-chs

10 Ideas to Fix Democracy
https://foreignpolicy.com/2022/01/07/10-ideas-fix-democracy/

Challenges to Liberal Democracy I: Reacting to Populism ...
https://ecpr.eu/Events/Event/PanelDetails/8522

Democracy in Crisis
https://freedomhouse.org/report/freedom-world/2018/democracy-crisis

The Past Decade and the Future of Governance and Democracy
https://www.bbvaopenmind.com/en/articles/the-past-decade-and-the-future-of-governance-and-democracy-populist-challenges-to-liberal-democracy/

Challenges to Democracy | Daniel Ziblatt - Scholars at Harvard
https://scholar.harvard.edu/dziblatt/challenges-democracy

Liberal democracy
https://en.wikipedia.org/wiki/Liberal_democracy

Can liberal democracy survive our fervor? Signs and ...
https://link.springer.com/article/10.1007/s12286-022-00525-9

Opinion | Is Liberal Democracy Dying?
https://www.nytimes.com/2022/09/28/opinion/italy-meloni-democracy-authoritarianism.html

Ten Challenges in Democracy Support – and How to ...
https://www.globalpolicyjournal.com/articles/global-governance/ten-challenges-democracy-support-and-how-overcome-them

Challenges to Democratic Governance in Developing ...
https://link.springer.com/book/10.1007/978-3-319-03143-9

Challenges for Democratic Governance
http://www.oas.org/sap/publications/1997/art/art_002_97_spa.pdf

Challenges to democracy | Inequality, Identity Politics ...
https://www.britannica.com/topic/challenges-to-democracy

Democratic Governance: Challenges and Obstacles in the ...
https://www.iri.org/news/democratic-governance-challenges-and-obstacles-in-the-midst-of-a-pandemic/

democratic-governance-effective-and-enduring-response. ...
https://www.idea.int/sites/default/files/publications/democratic-governance-effective-and-enduring-response.pdf

Challenges to Democracy Building
https://www.corteidh.or.cr/tablas/25040.pdf

Solutions for Democratic Decay
https://www.dissentmagazine.org/article/solutions-for-democratic-decay/

THE CHALLENGE OF DEMOCRATIC GOVERNANCE IN ...
https://papers.ssrn.com/sol3/Delivery.cfm/SSRN_ID2381312_code1520621.pdf?abstractid=2381312&mirid=1

Cracks within democracies & the declining liberal practices
https://www.prosperity.com/feed/cracks-within-democracies-declining-liberal-practices

The Enduring Vulnerability of Liberal Democracy
https://www.journalofdemocracy.org/articles/the-enduring-vulnerability-of-liberal-democracy/

2. What Crisis of Liberal Democracy?
https://www.chathamhouse.org/2020/03/future-democracy-europe/2-what-crisis-liberal-democracy

Figure A4: Examples of generated solutions after being evolved for 15 generations by the genetic algorithm.

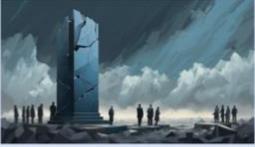

Figure A5: Examples of pros and cons generated for each solution.

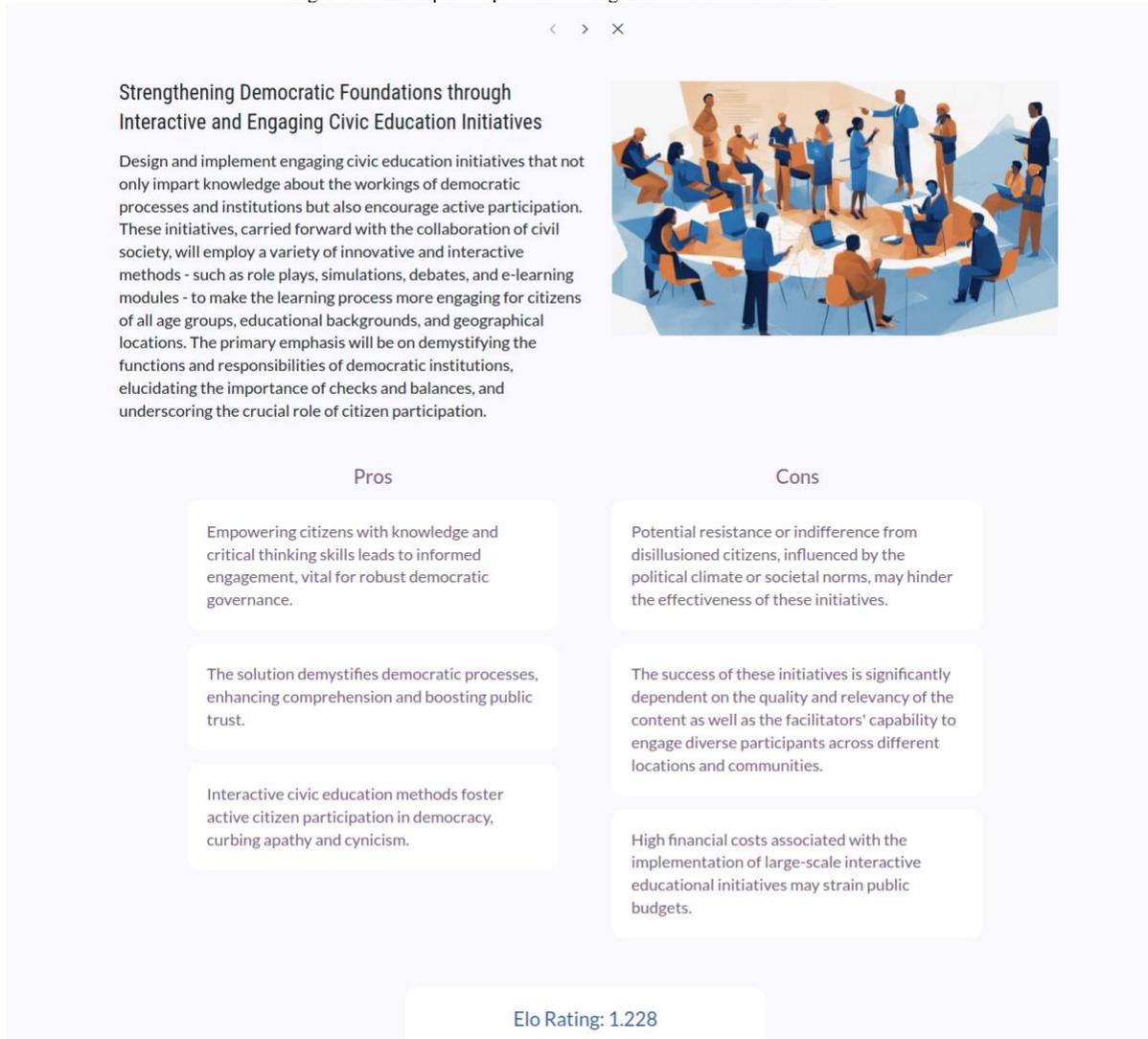

Figure A6: Examples of solutions undergoing generations of evolution.

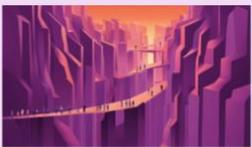

Figure A7: Examples of pros and cons for evolved solution.

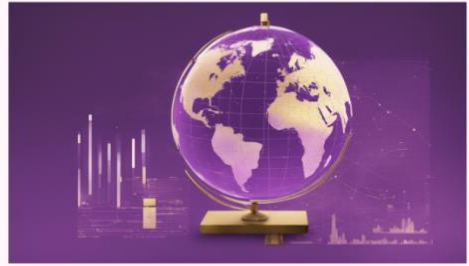

Figure A8: Evolved solutions after 30 generations.

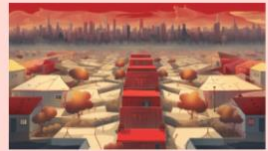
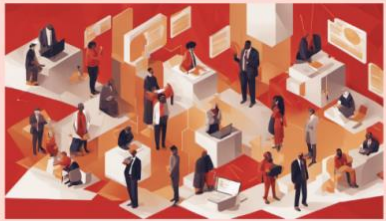
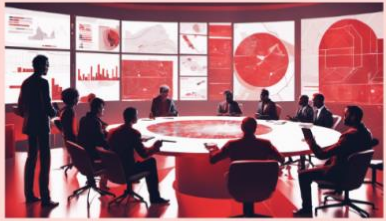

Figure A9: Solutions evolving further into policies.

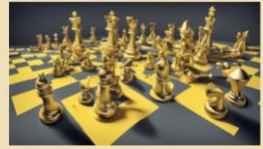
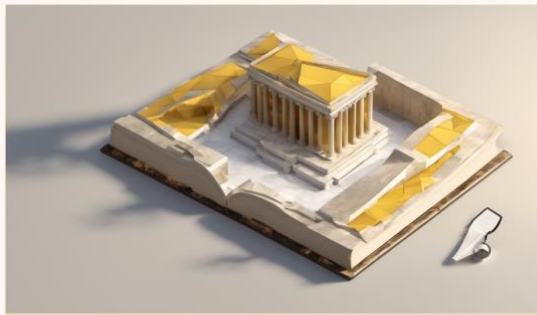
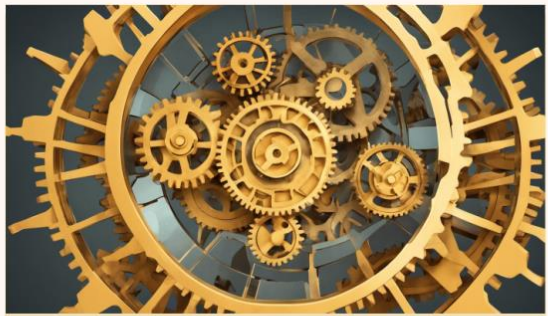

Figure A10: Policies beginning to undergo further evolution.

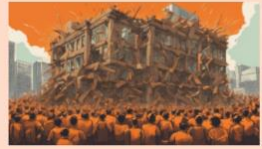
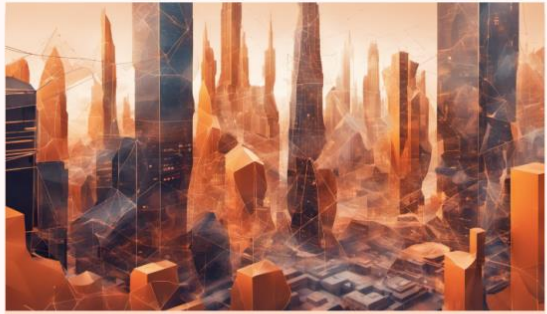
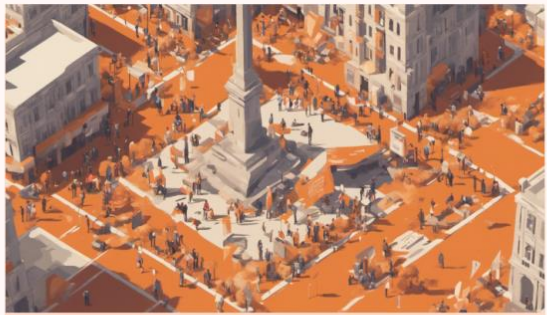

Figure A11: Example of evidence and analysis for one policy recommendation.

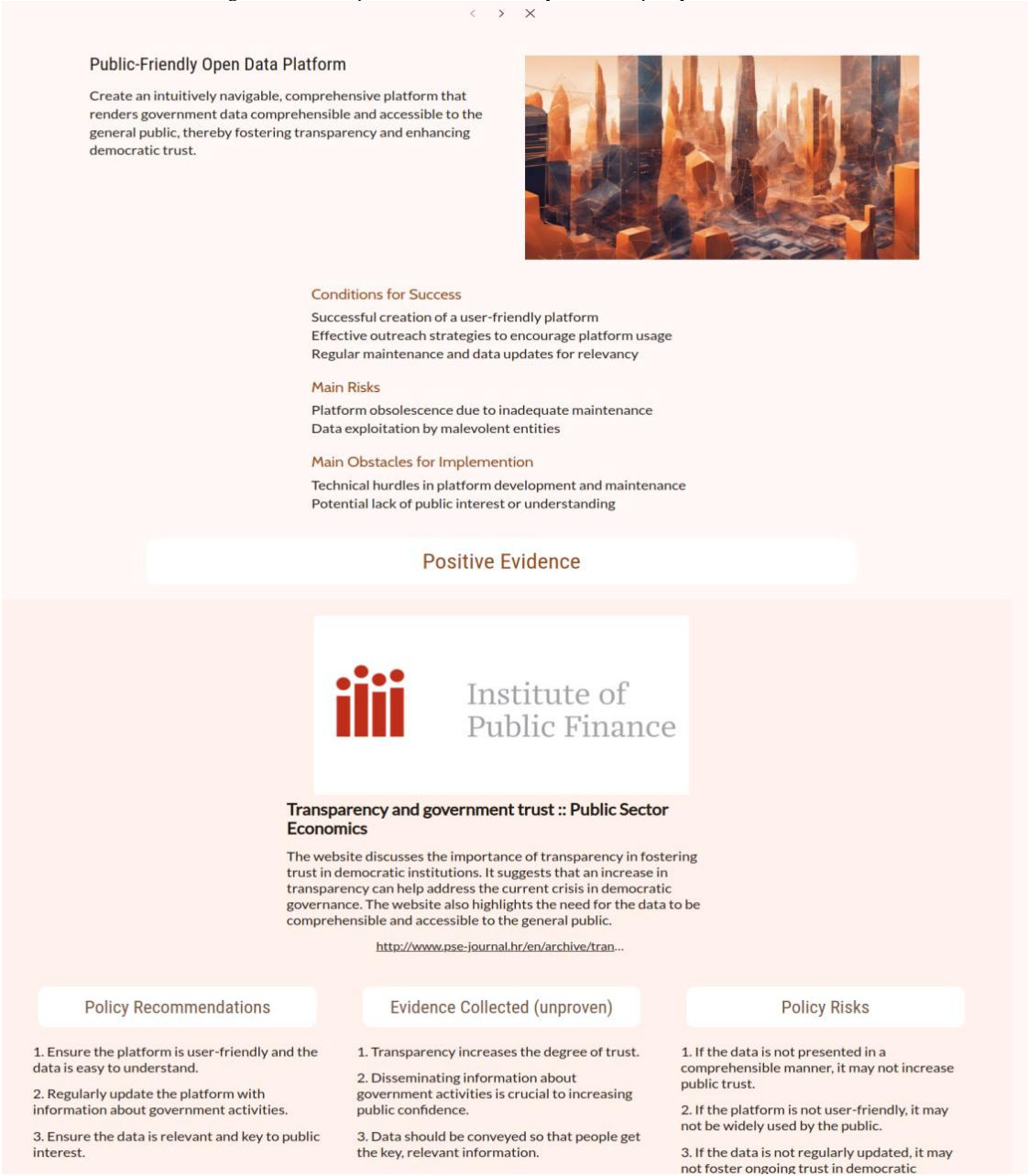

Figure A12: Example of Policy Synth's analysis of a specific solution.

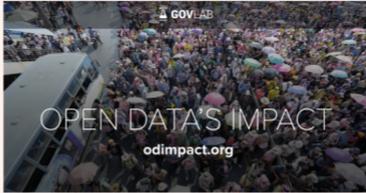

Figure A13: An additional example of Policy Synth's analysis of a specific solution.

## Case Studies

**GOVLAB**

The website presents a detailed analysis of 19 open data projects from around the world, highlighting their impacts and the conditions that enable their success. It concludes with a set of recommendations for those considering implementing open data projects.

https://thegovlab.org/static/files/publicati...

### Policy Recommendations

1. Ensure partnerships and collaborations among various organizations.

2. Establish an open data public infrastructure for regular release of impactful data.

3. Implement clear open data policies, including performance metrics.

More (14)

### Evidence Collected (unproven)

1. Open data projects can improve government accountability and efficiency.

2. Open data projects can empower citizens by facilitating informed decision-making and enabling social mobilization.

3. Open data projects can create new economic opportunities.

More (15)

### Policy Risks

1. Potential lack of readiness, especially in terms of technical and human capacity.

2. Risk of projects being unresponsive or inflexible to user or citizen needs.

3. Potential inadequate protections for privacy or security.

More (8)

### Pros for Policy from source

1. Increased government accountability and efficiency.

2. Empowered citizens through informed decision-making and social mobilization.

3. Creation of new economic opportunities.

More (11)

### Cons for Policy from source

1. Potential lack of readiness, especially in terms of technical and human capacity.

2. Risk of projects being unresponsive or inflexible to user or citizen needs.

3. Potential inadequate protections for privacy or security.

More (5)

### Academic Sources

1. GovLab

2. The GovLab

3. Open Data Partnership for Development (ODP4D)

More (2)

**Policy Synth Example 2: Modernizing Policymaking**

**Problem Statement used:** In modern democracies, government policymaking processes are increasingly struggling to keep pace with the rapidly evolving dynamics of society. This struggle is evident in the delayed implementation of policies, a growing misalignment between governmental actions and public needs, and a general inflexibility in adapting to swift societal changes. As the pace of change in society continues to accelerate, these concern about their current effectiveness and future readiness challenges within the policymaking framework are becoming more evident, highlighting a significant

Figure A14: Top sub problems identified for Modernizing Policymaking topic.

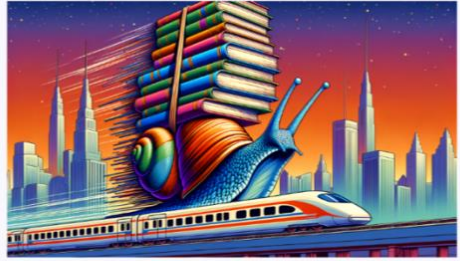

In modern democracies, government policymaking processes are increasingly struggling to keep pace with the rapidly evolving dynamics of society. This struggle is evident in the delayed implementation of policies, a growing misalignment between governmental actions and public needs, and a general inflexibility in adapting to swift societal changes. As the pace of change in society continues to accelerate, these challenges within the policymaking framework are becoming more evident, highlighting a significant concern about their current effectiveness and future readiness

Sub Problems and Solutions

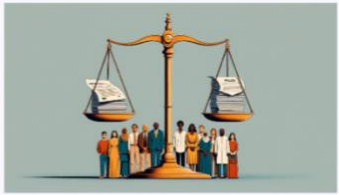

**Lack of Norms-Policy Research**

Insufficient research on policy impact on social norms hinders effective policymaking.

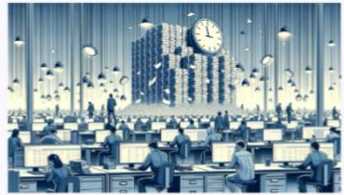

**Information Processing Inefficiencies**

Government bodies struggle to effectively process and analyze the vast amounts of information available.

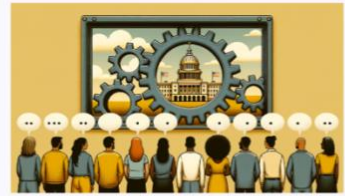

**Ineffective Citizen Participation**

There is a disconnect between the policy development process and citizen input.

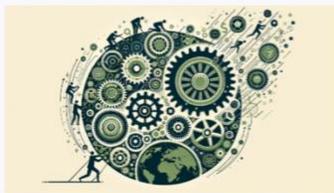

**Economic Shifts**

Global economic shifts alter societal dynamics faster than policies can adjust.

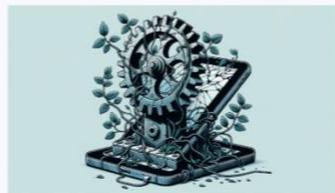

**Systemic Inflexibility**

Legacy systems in government are slow to adapt.

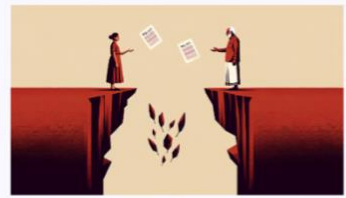

**Generational Perception Gaps**

Generational gaps in policy perception affect policy uptake.

Figure A15: Examples of generated solutions after being evolved for 5 generations by the genetic algorithm:

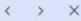

Figure A16: Examples of pros and cons and evolutionary tree generated for one solution.

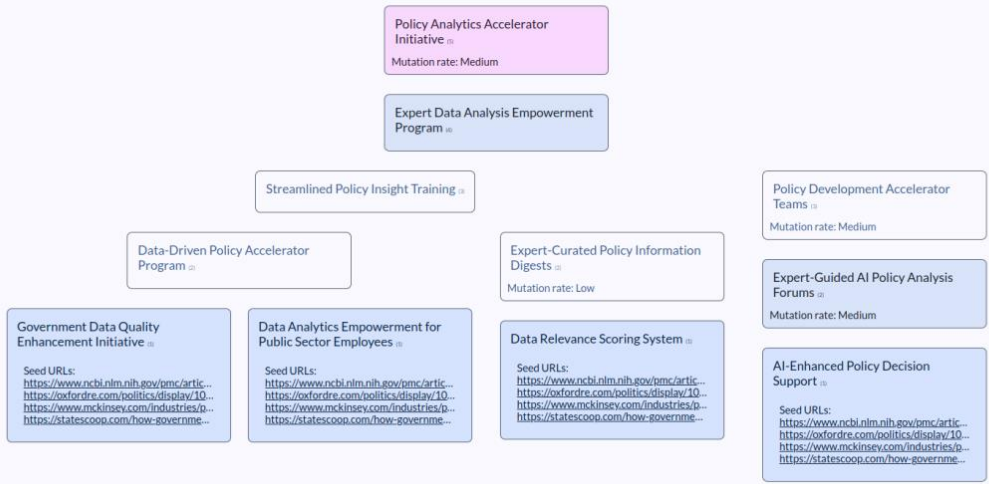

Figure A17: Examples of two more solutions after five generations of evolution.

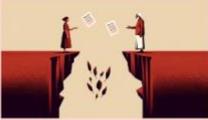

Figure A17: Examples of pros and cons and evolutionary tree generated for another solution.

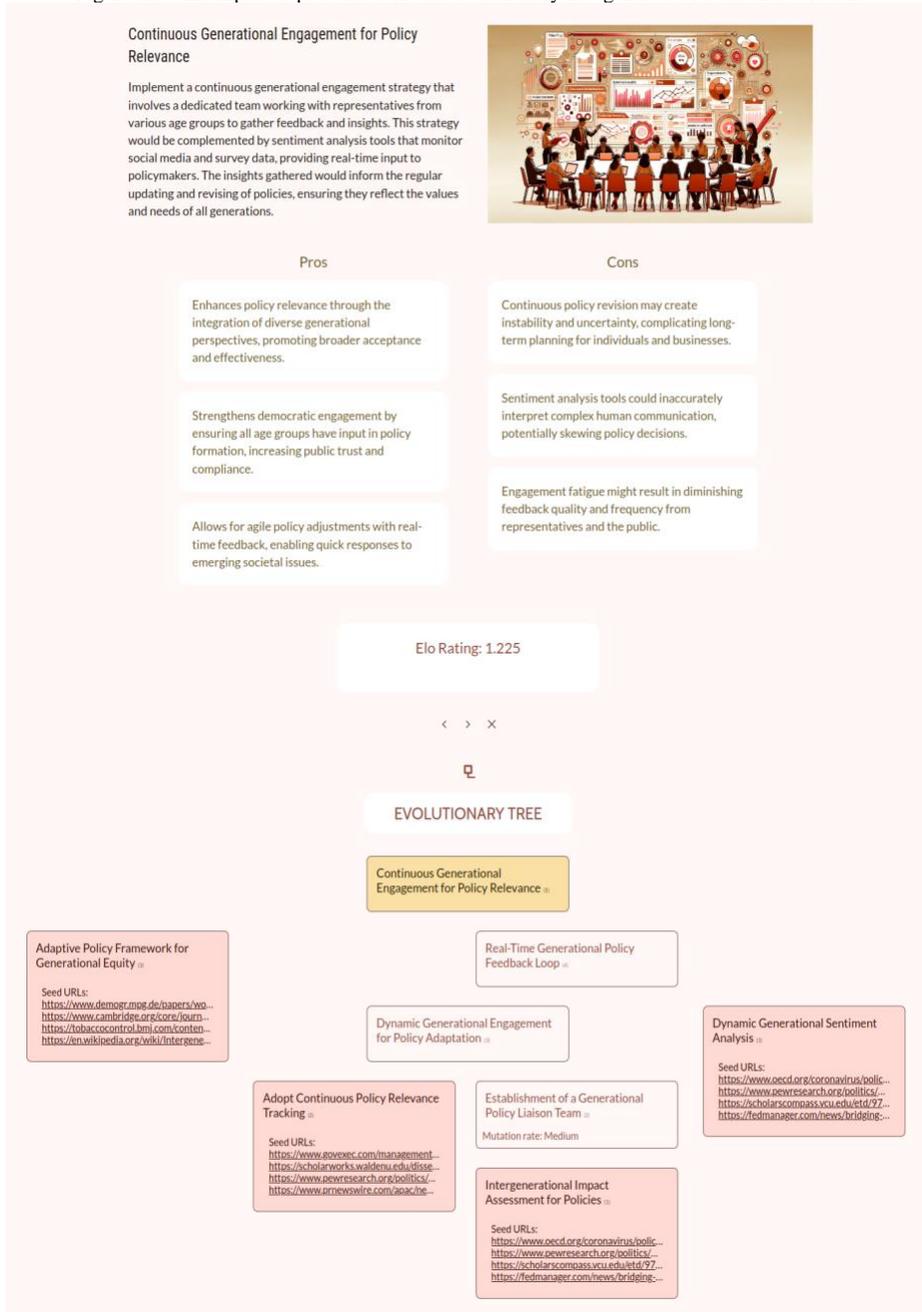

**Example Policy Synth System 1 Agent Prompts**

*Example Pairwise Voting System 1 Agent/Prompt*

System message: You're an expert in evaluating and ranking solution components to problems.

    Instructions:
    1. Analyze a problem and two solution components, labeled "Solution Component One" and "Solution Component Two"
    2. Determine which is more important and practical.
Important Instructions:
   1. Keep in mind that the solution components will be implemented by philanthropic organizations in partnership with civil society organizations, community-based organizations, and legal advocacy groups.

Always output your decision as "One", "Two" or "Neither. No explanation is necessary.

Think step by step.
User message

**Problem:**

Persistent Political Instability
Consistent instability due to frequent changes in government or policy introduces a climate of uncertainty and inconsistency.

Such instability can potentially harm economic performance, provision of public services, and the nation's international standing, thereby undermining public confidence in democratic governance.

**Solutions to assess:**

**Solution Component One:**
Revitalizing Democracy through Education, Modernization, and Digital Inclusion

This approach to democratic revitalization encompasses a comprehensive effort towards enhanced civic education, progressive institutional reforms, and the pervasive use of digital solutions. Through cultivating a well-informed citizenry via civic education programs, we can foster more effective participation in democratic processes. Institutions should undergo strategic modernization to offer efficient services, uphold a performance-based culture, and establish robust anti-corruption safeguards. The implementation of digital tools, including e-governance platforms and AI-enabled applications, will significantly streamline bureaucratic functions, elevate public service standards, encourage wider participation, and improve transparency in real-time. This methodology aims to restore public faith, augment institutional performance, and reinforce the pillars of democratic governance.

**Top pro for solution component one:**
This approach integrates enhanced civic education, institutional modernization, and digital tools to promote informed citizen participation, improve government efficiency, and increase transparency, thereby strengthening democratic processes.

**Top con for solution component one:**
The reliance on digital solutions risks exacerbating the digital divide, potentially marginalizing those without access to technology or digital literacy, thus undermining equitable democratic participation.

**Solution Component Two:**
Boosting Public Trust through Participatory Budgeting

Institute participatory budgeting processes at local government levels, enabling citizens to have a direct say in the allocation of a segment of the public budget. This approach enhances transparency and accountability while empowering citizens through active involvement in decision-making, thereby bolstering their trust in democratic institutions.

**Top pro for solution component two:**
Allowing citizens to directly participate in budget allocation increases the transparency of governmental financial decisions. This can reduce corruption and misuse of funds, as public scrutiny is heightened.

**Top con for solution component two:**
There's a risk that only a small, possibly more privileged segment of the community will participate, leading to decisions that don't represent the wider population.

**The more important solution component is:**

Answer
**One**

*Example GA S1 Agent: Random mutation*

**System message:**
As an AI expert specializing in genetic algorithms, your task is to mutate the following solution.

**Instructions:**
1. Implement mutations corresponding to a high mutation rate.
2. Mutation can involve introducing new attributes, modifying existing ones, or removing less important ones.

3. Ensure the mutation is creative, meaningful, and continues to offer a viable solution to the presented problem.

4. Avoid referring to your output as "the merged solution" or "the mutated solution". Instead, present it as a standalone solution.

Important Instructions (override the previous instructions if needed):
1. Never create solutions in the form of frameworks or holistic approaches
2. Solutions should include only one core idea.
3. The solution title should indicate the benefits or results of implementing the solution.
4. Remember that the main facilitator for implementation will be civil society working with governments.
5. Frame solutions with the intention of convincing politicians and governments to put them into action.

Always format your mutated solution in the following JSON structure:
{ title: string, description: string, mainBenefitOfSolution: string, mainObstacleToSolutionAdoption: string }.

Do not introduce any new JSON properties.
Think step by step.

User message:

**Problem Statement:**
Liberal democracies are grappling with an overarching issue of declining effectiveness and stability, which is leading to widespread citizen dissatisfaction and threatening the fundamental principles of democratic governance.

**Sub Problem:**
Prevalence of Corruption

Widespread corruption and misuse of power erode the effectiveness and credibility of democratic institutions.

**Top Affected Entities:**

Citizens

**Negative Effects:**
Perceived or actual corruption escalates public dissatisfaction which can lead to social unrest and political instability.

Loss of public trust in democratic governance due to corruption discourages citizen participation in democratic processes.

*Democratic Institutions*

**Negative Effects:**

Erosion of credibility and effectiveness due to rampant corruption hampers the functionality of institutions and reduces public faith.

Distortion of their intended function due to the misuse of power leads to inefficient operation and suboptimal outcomes.

*Political Leadership*

**Negative Effects:**
Increased public scrutiny and loss of trust due to corruption can lead to a decline in the political capital for leaders.

Involvement in corrupt practices can lead to legal consequences, damaging both the personal and professional life of the leaders.

**Solution to mutate:**
```
  {
    "title": "Encourage Citizen Oversight through Digital Transparency Reforms",
    "description": "Promote digital transparency reforms to mitigate corruption. Ensure these reforms protect media and internet freedom. Leverage open government data to enhance transparency and accountability, and encourage citizens to use this data to monitor government activities and hold public servants accountable.",
    "mainBenefitOfSolution": "This solution can help limit corruption by promoting a culture of transparency and accountability through citizen involvement.",
    "mainObstacleToSolutionAdoption": "Potential resistance from public servants and a potential lack of technical expertise to implement these reforms may pose challenges."
  },
```

Generate and output JSON for the mutated solution below:

**Answer:**

```
{
    "title": "Implement Whistleblower Protection and Reward Systems",
    "description": "Introduce a comprehensive whistleblower protection program that safeguards individuals who report corrupt practices. This program should include legal protections against retaliation, anonymity options for reporters, and a reward system that compensates whistleblowers for their risks and contributions to exposing corruption. By creating a secure environment for whistleblowers, this approach encourages insiders to come forward with information on corruption without fear of repercussions. This initiative should
```

```
be widely publicized to ensure potential whistleblowers are aware of the protections and
incentives available to them.",

    "mainBenefitOfSolution": "This solution empowers individuals within organizations to act
against corruption by providing them with a safe channel to expose unethical practices,
thereby enhancing the integrity of democratic institutions.",

    "mainObstacleToSolutionAdoption": "The main challenge could be political resistance from
those who might view such protections as a threat to their interests, alongside the logistical
and financial considerations involved in setting up and maintaining the whistleblower
protection and reward systems."
}
```

**DETAILED RESULTS OF HUMAN V. AI COMPARATIVE TEST**

Here is all primary data for the human v. AI comparative test.

The full source code to run the automated GPT-4 rating comparisons is available in the Policy Synth GitHub archive.

Election Violence

External link: https://docs.google.com/spreadsheets/d/1BN6yCQXRsOoE49u0JfiDTbwrtxwGkB9V3P5CJssWMak

Misuse of Legal system

External link: https://docs.google.com/spreadsheets/d/1GuL29z6epHwuiQbFgOfOeLbD9B932VorVhLunmJzhcE/

**RUNNING POLICY SYNTH**

Information about Policy Synth, documentation and the agent software itself is available on GitHub here https://github.com/CitizensFoundation/policy-synth